\documentclass{article} 
\usepackage{iclr2024_conference,times}

\usepackage{amsmath,amsfonts,bm}

\def\eqref#1{equation~\ref{#1}}

\def\1{\bm{1}}

\DeclareMathAlphabet{\mathsfit}{\encodingdefault}{\sfdefault}{m}{sl}
\SetMathAlphabet{\mathsfit}{bold}{\encodingdefault}{\sfdefault}{bx}{n}

\usepackage{hyperref}
\usepackage{url}
\usepackage{graphicx}
\usepackage{booktabs}
\usepackage{tabularx}
\usepackage{enumitem}
\usepackage{setspace}

\title{Fine-tuning protein language models with deep mutational
 scanning improves variant \\ effect prediction }

\author{Aleix Lafita$^{1,\dagger, *}$ 
\And
Ferran Gonzalez$^{2, \dagger, *}$ 
\And
Mahmoud Hossam$^{2}$ 
\And
Paul Smyth$^{2}$
\And
Jacob Deasy$^{2}$
\And
Ari Allyn-Feuer$^{2}$
\And
Daniel Seaton$^{1}$
\And
Stephen Young$^{2}$
\\
\AND
\\
$^{1}$ GSK, Human Genetics and Genomics \\
$^{2}$ GSK, Artificial Intelligence and Machine Learning
\\ 
$^{\dagger}$ equal contribution \\
$^{*}$\texttt{\{aleix.x.lafita,ferran.x.gonzalez\}@gsk.com}
 \\
}

\iclrfinalcopy 
\begin{document}

\maketitle

\begin{abstract}

Protein Language Models (PLMs) have emerged as performant and scalable tools for predicting the functional impact and clinical significance of protein-coding variants, but they still lag experimental accuracy. Here, we present a novel fine-tuning approach to improve the performance of PLMs with experimental maps of variant effects from Deep Mutational Scanning (DMS) assays using a Normalised Log-odds Ratio (NLR) head. We find consistent improvements in a held-out protein test set, and on independent DMS and clinical variant annotation benchmarks from ProteinGym and ClinVar. These findings demonstrate that DMS is a promising source of sequence diversity and supervised training data for improving the performance of PLMs for variant effect prediction. 
\end{abstract}

\begin{spacing}{0.98}
\section{Introduction}

The pace of discovery of new human genetic variants is rapidly increasing, led by genome sequencing of large human cohorts \citep{lek2016analysis, backman2021exome}. However, the functional characterisation of these human variants has not scaled at the same pace, limiting their impact for clinical diagnosis and drug target discovery, and ultimately hampering our ability to understand and treat human diseases \citep{landrum2018clinvar}. Missense variants are the most common type of coding variant, causing single amino acid changes in proteins that can have a wide range of consequences, from severe disease-causing protein function disruptions to no significant effect. Predicting the impact of missense mutations on protein function and the downstream clinical consequences remains a crucial challenge \citep{karczewski2020mutational}. 

Over the last few years, advancements in deep learning have led to significant improvements for tackling the missense variant effect prediction challenge \citep{frazer2021disease, cheng2023accurate, gao2023landscape}. Protein Language Models (PLMs) have demonstrated state-of-the-art (SOTA) performance in accuracy and generalisability at various protein variant effect prediction tasks \citep{brandes2023genome, lin2023evolutionary, rives2021biological}, but there are still gaps in their performance for clinical variant classification and correlation with experimental assays \citep{livesey2023updated}.

In this study, we improve the performance of PLMs for variant effect prediction using experimental scores from Deep Mutational Scanning (DMS). We first introduce a rescaling and normalisation pipeline to integrate DMS assays from multiple proteins into a common functional scale. We then present a novel lightweight fine-tuning approach for PLMs named Normalised Log-odds Ratio (NLR) – that can efficiently learn from DMS data by adding parameter-free layers on top of the language modelling head of PLMs. We finally evaluate the performance improvements of our approach on held-out test proteins and independent DMS and clinical annotation benchmarks, while ensuring low sequence similarity between training and evaluation proteins to assess model generalisation. 

\end{spacing}
\section{Background}

\textbf{Deep Mutational Scanning (DMS)} is an experimental technique that leverages high-throughput DNA sequencing and fitness selection assays to exhaustively measure the effect of variants in a protein region \citep{fowler2014deep}, providing comprehensive maps of protein variant effects that can be used to understand the clinical relevance of human genetic variants \citep{findlay2018accurate, radford2023saturation}. Recognising the rapidly expanding volume of DMS data and the challenges with data compilation and reproducibility, the Atlas of Variant Effects (AVE) alliance \citep{fowler2021atlas} developed MaveDB, an open-source repository of DMS assays \citep{esposito2019mavedb, rubin2021mavedb}. More recently, ProteinGym has emerged as an independent collection of manually curated DMS assays, providing a standardised framework for benchmarking protein fitness prediction and design \citep{notin2023proteingym}. Despite the wide adoption of DMS datasets for benchmarking variant effect prediction models \citep{livesey2023updated}, one of the remaining challenges in combining DMS scores across assays and proteins is that their scale is highly dependent on experimental methods and selection assays, requiring rescaling and normalisation \citep{dunham2021exploring}.

\textbf{Protein Language Models (PLMs)} are pre-trained on large corpora of naturally evolved protein sequences using self-supervision and have shown great promise predicting the impact of missense variants without additional supervision (i.e.\,zero-shot) \citep{meier2021language}. The Evolutionary Scale Modelling (ESM) family of PLMs \citep{rives2021biological,meier2021language,lin2023evolutionary} pre-trained with masked language modelling on the UniRef database \citep{uniprot2023uniprot} demonstrated the ability to encode functional and structural patterns crucial for variant effect and structural predictions. More recently, AlphaMissense \citep{cheng2023accurate}, pre-trained on the Protein Data Bank (PDB) \citep{wwpdb2019protein} and fine-tuned on population frequency data, achieved SOTA results in predicting the clinical pathogenicity of human missense variants. The success of PLMs in zero-shot variant effect predictions has led to the use of fine-tuning approaches to improve performance on specific tasks, including protein stability \citep{umerenkov2023prostata}, pathogenicity \citep{lin2023varipred}, protein-protein interactions \citep{sledzieski2023democratizing}, secondary structure and sub-cellular location \citep{schmirler2023fine}, and protein fitness \citep{rives2021biological,hsu2022learning,jagota2023cross}.

\section{Methods}
\label{gen_inst}

\begin{figure}[h]
   \begin{center}
   \includegraphics[width=0.9\textwidth]{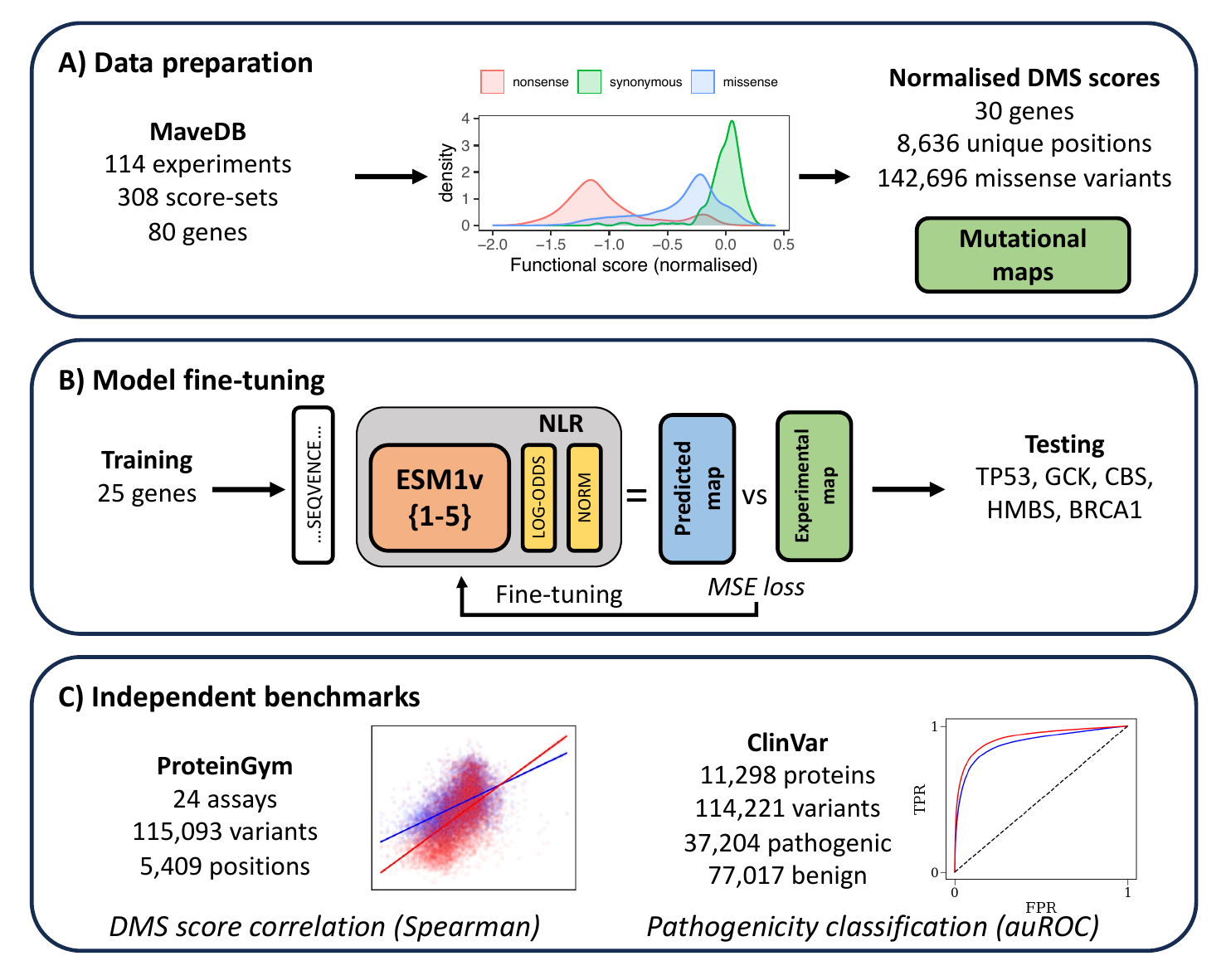}
\end{center}
   \caption{
      Methods overview. 
   A) Preparation of normalised DMS functional scores from a subset of MaveDB experiments. The mean scores of synonymous and nonsense variants are used to create a common scale across assays and proteins.
   B) Fine-tuning pipeline for ESM-1v models using the Normalised Log-odds Ratio (NLR) head.
   C) Performance evaluation on two independent benchmarks: DMS assays from ProteinGym, and pathogenic and benign missense variants from ClinVar.}
     \label{fig:summary}
\end{figure}

\subsection{Training and evaluation datasets}

\paragraph*{Normalisation of DMS datasets from MaveDB.} We downloaded a total of 308 score-sets from 113 experiments in MaveDB (\href{https://www.mavedb.org}{https://www.mavedb.org}) in July 2023. We selected experiments of type \textit{Protein coding}, manually mapped targets to 80 unique gene names and reassigned variants to UniProt sequence positions. We then categorised variants into \textit{nonsense}, \textit{missense} and \textit{synonymous} types using the \textit{hgvs\_pro} column, and filtered out indels (insertions and deletions) and multiple amino acid variants. We filtered out viral proteins and selected datasets with at least one synonymous, one nonsense, and over 50 missense variants, resulting in 103 datasets for 30 proteins (Table \ref{tab:protlist}). To normalise functional scores across all datasets, we converted them to log-scales and rescaled the distribution so that the mean score of synonymous variants ($S_{syn}$) was 0 and the mean score for nonsense variants ($S_{nonsense}$) was -1, using equation \ref{eq:experimental} (Figure \ref{fig:summary}A, Figure \ref{fig:mavedb}). Values were further capped in the [-2, 2] range to limit outliers. 
\begin{equation}
   S_{norm} = \frac{S_{raw} - mean(S_{syn})}{mean(S_{syn}) - mean(S_{nonsense})} \label{eq:experimental}
\end{equation}

We then selected scores for missense variants and aggregated assays for each protein. In the case of duplicated scores for the same protein variant in multiple assays, we calculated the mean score. The final dataset contained 142,696 missense variants covering 8,636 unique protein positions in the 30 genes. We clustered the 30 protein sequences using MMseqs2 (version 14) \citep{steinegger2017mmseqs2} with 20\% coverage and 20\% sequence identity thresholds, yielding 29 unique clusters (only CCR5 and CXCR4 were clustered together). We selected the 5 proteins with most ClinVar labels (TP53, GCK, CBS, HMBS, and BRCA1) for model testing (Table \ref{tab:protlist}). The remaining clusters were used to select variants in 25 proteins for model training and cross-validation. 

\paragraph*{Benchmarking DMS datasets from ProteinGym.} ProteinGym provides functional scores for two types of variants: amino acid substitutions and indels. Since ProteinGym does not provide scores for synonymous or nonsense variants, the normalisation approach needed for model training that we propose here was not possible. However, ProteinGym assays can still be used for benchmarking using correlation metrics. We downloaded ProteinGym substitution scores for a subset of 43 DMS assays with open licenses from the ProteinGym website (\href{https://proteingym.org/}{https://proteingym.org}) in November 2023. We then removed 19 assays with sequence similarity to our 25 MaveDB training proteins using the same procedure described above and selected variants with single amino acid substitutions, resulting in a final benchmark dataset of 24 assays and 115,093 missense variants (Figure \ref{fig:summary}C).
\paragraph*{ClinVar pathogenic and benign variants.} ClinVar is a public archive of human genetic variants and interpretations of their significance to disease \citep{landrum2018clinvar}. We downloaded the complete set of 862,666 variant annotations in the \textit{variant\_summary.txt} file (version 2023\_04) from ClinVar's FTP website (\href{https://ftp.ncbi.nlm.nih.gov/pub/clinvar/}{https://ftp.ncbi.nlm.nih.gov/pub/clinvar}). We then selected missense variants mapped to protein transcripts  (\textit{HGVSp} column), with at least one reviewer star (\textit{ReviewStatus} column), and with clinical significance (\textit{CLNSIG} column) labels as \textit{Benign} and \textit{Likely benign} (considered as \textit{benign} in this study), and variants labelled as \textit{Pathogenic} and \textit{Likely pathogenic} (considered as \textit{pathogenic}), similar to previous work \citep{cheng2023accurate, lin2023varipred}. We then removed 430 proteins with sequence similarity to our 25 MaveDB training proteins using the same procedure described above. The final ClinVar dataset contains 114,221 missense variants in 11,298 proteins (Figure \ref{fig:summary}C). To evaluate per-protein performance of our models, we constructed a balanced dataset by selecting proteins with at least 10 benign and 10 pathogenic labels, similar to \citet{cheng2023accurate}, resulting in 37,142 variant annotations for 361 proteins. 

\begin{spacing}{0.95}
\subsection{Normalised Log-odds Ratio}

We present a Normalised Log-odds Ratio (NLR) framework (Figure \ref{fig:summary}B), which allows efficient fine-tuning on DMS data by adding parameter-free layers on top of PLMs. The key components of the NLR architecture are shown in Figure \ref{fig:finetuning} and include:

\begin{enumerate}[leftmargin=*]
  
   \item[1.] \textbf{Training instances}:  Individual amino acid substitutions with corresponding DMS scores. 
   \item[2.] \textbf{Variant representation}: For each training instance we provide the wildtype protein as input, allowing efficient computation of all missense variant effects within a protein in a single forward pass during inference. We also experimented masking input tokens to encourage model regularisation, but simply using the raw wildtype sequence outperformed the masking strategies explored (Appendix \ref{subsec:masking}).
   \item[3.] \textbf{Encoder}: We chose ESM as our baseline models and initialised the encoder weights from their corresponding checkpoints. In contrast to most fine-tuning approaches \citep{schmirler2023fine, lin2023varipred}, NLR fine-tunes both the pre-trained transformer encoder blocks and the masked language modelling head. We used ESM-1v (esm1v\_t33\_650M\_UR90S\_[1-5]) for most experiments but also benchmarked NLR when using ESM-2 (esm2\_t33\_650M\_UR50D) and ESM-1b (esm1b\_t33\_650M\_UR50S) as pre-trained models. 
   \item[4.1] \textbf{NLR - Log-odds Ratio computation}: NLR follows the inference approach of existing zero-shot variant effect prediction methods, scoring all possible amino acid substitutions in a sequence by calculating the log-odds ratio between reference and alternate amino acid probabilities at each mutated position (e.g. \citet{meier2021language, cheng2023accurate}). However, NLR performs this computation during both, fine-tuning and inference phases. This is facilitated by the NLR head which, given a wildtype sequence, computes the matrix of log-odds ratios in a differentiable form. As visualised in Figure \ref{fig:finetuning}, this computation results in a matrix of sequence length by vocabulary size after each forward pass, providing a score for all possible amino acid substitutions.
   \item[4.2] \textbf{NLR - Normalisation}: To address the distinct scale between log-odds ratios and DMS scores, NLR applies a normalisation layer to each log-odds ratio matrix during training. After exploring the effect of different approaches, such as instance normalisation \citep{ulyanov2016instance}, min-max normalisation between -2 and 2 proved superior to other methods tested (Appendix \ref{fig:finetuning}). 
   \item[5.] \textbf{Loss and output} During training, the predicted score for a specific variant is retrieved by indexing the corresponding position in the normalised log-odds ratio matrix. This score is then compared to the ground-truth DMS score to compute the mean squared error (MSE) loss for backpropagation. 

\end{enumerate}

\paragraph*{Training and evaluation.} We used the subset of 25 MaveDB proteins for model training and explored hyperparameters and architecture choices with 5-fold cross-validation due to varied zero-shot correlation across proteins (Appendix \ref{subsec:finetuning}). We fine-tuned all model parameters as early experiments exhibited larger performance gains than freezing transformer encoder blocks (Appendix \ref{subsec:freezing}). After conducting cross-validation experiments, the final models were trained using the entire dataset, consisting of DMS data for 109,215 variants from 25 proteins. Final runs were trained for 2,000 optimization steps to prevent overfitting, which was observed during the cross-validation experiments (Appendix \ref{subsec:finetuning}). Since ESM-1v is a five-model ensemble, five independent runs were performed, each starting with a different ESM-1v checkpoint. At inference time, new variants were scored by averaging the log-odds ratio scores across the five model predictions.

 \section{Results}
 
\begin{figure}[h]
   \begin{center}
   \includegraphics[width=0.97\textwidth]{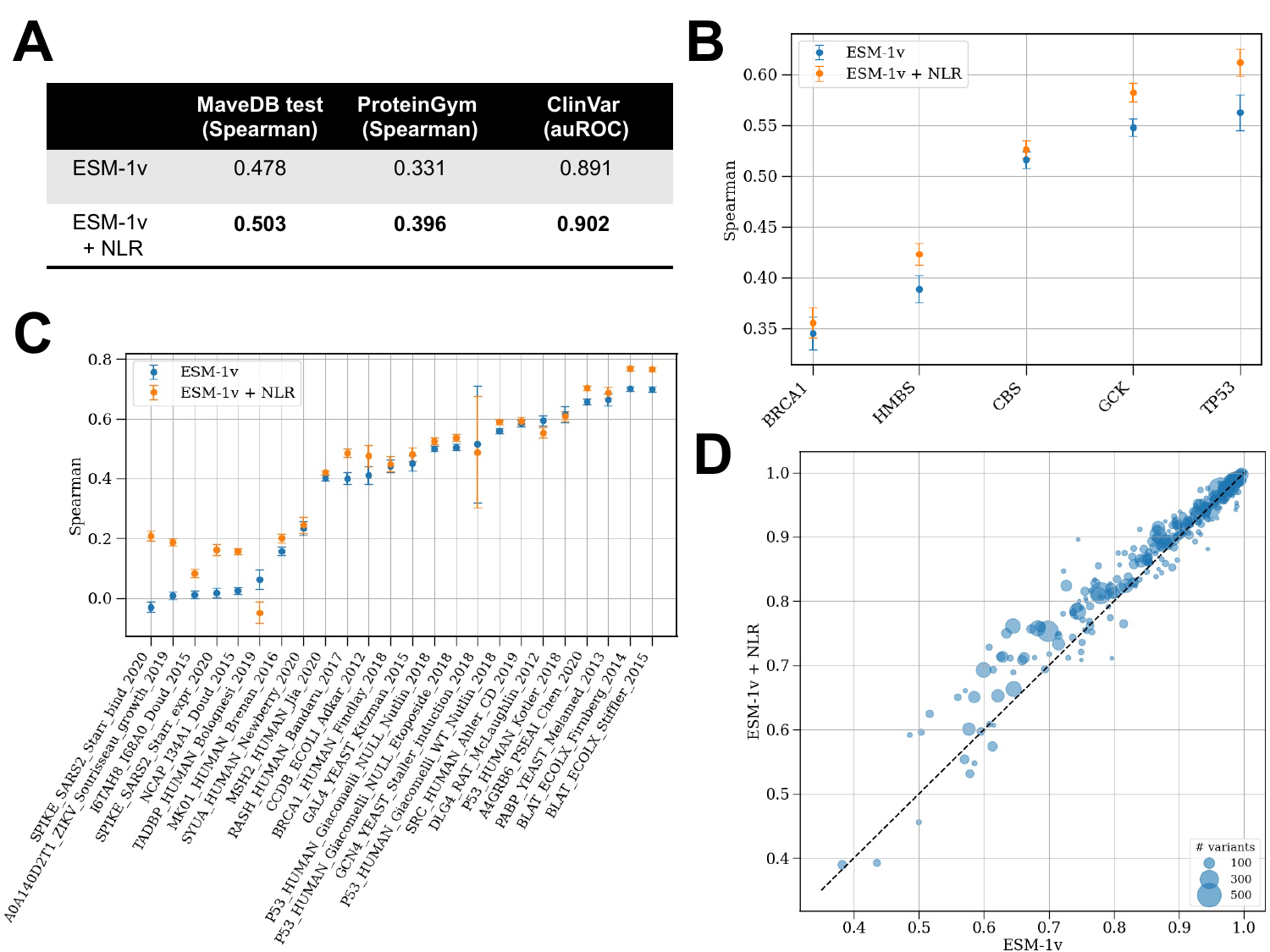}
\end{center}
   \caption{
      Results after NLR fine-tuning of ESM-1v models across benchmarks.
      A) Performance in the five MaveDB test proteins. ProteinGym DMS assays and ClinVar pathogenic variants.
       B) Spearman correlation in MaveDB test proteins. Mean $\pm$ standard deviation (std) of 50 bootstrapped samples. 
       C) Spearman correlation in ProteinGym DMS assays. Mean $\pm$ std of 50 bootstrapped samples. 
       D) Per-protein auROC for ClinVar proteins with over 10 benign and 10 pathogenic variants. 
   }
   \label{fig:results}
\end{figure}

 We ran inference with pre-trained and NLR fine-tuned models on the five MaveDB test proteins, the 24 ProteinGym DMS assays and the 114,221 ClinVar pathogenic and benign variants. As shown in Figure \ref{fig:results}A, NLR fine-tuning of ESM-1v improves the performance of missense variant effect predictions across all benchmarks. 
 
 \paragraph*{Improved MaveDB DMS predictions.} Our MaveDB test set was composed of five human genes (TP53, GCK, CBS, HMBS and BRCA1) with associated DMS data, which went through the filtering and standardisation procedure presented in this study. 
 From the results in Figure \ref{fig:results}A, we find that NLR fine-tuning of ESM-1v ensemble improves the micro-averaged\footnote{Micro-averaging of metrics applies weights per protein proportional to their relative number of variants, while macro-averaging applies equal weight per protein.} Spearman correlation across proteins in our MaveDB test set  from 0.478 to 0.503 (+5.2\%, \autoref{fig:results}A). 
 In Table \ref{tab:esm1vs}, the performance is stratified by each of the five ESM-1v model checkpoints, displaying improvements from 5.4\% to 9.4\% across all model versions. These results indicate that the fine-tuning improvements on individual model checkpoints are larger than the ensemble of ESM-1v models. 
 Figure \ref{fig:results}B displays the performance of ESM-1v per protein, exhibiting improvements after fine-tuning on all five proteins in the test set. We also assessed the impact of the number of proteins used for training (Appendix \ref{subsec:datascale}). We observed an upward trend in model performance with increasing  number of  training proteins, indicating that NLR fine-tuning can scale with more available DMS data.

\paragraph*{Improved ProteinGym DMS predictions.} Our ProteinGym benchmark was composed of 24 DMS assays which differed from the five proteins in the MaveDB test set in two ways: they included DMS data for non-human proteins and we did not apply our MaveDB pre-processing pipeline. From the results in Figure 2A we find that NLR fine-tuning of ESM-1v improves the average Spearman correlation across assays in ProteinGym from 0.331 to 0.396 (+19.6\%, \autoref{fig:results}A). Figure \ref{fig:results}C reveals that this improvement extends to nearly all DMS studies, both human and non-human. Notably, the improvements were larger for viral proteins, which were excluded from the training set and had the lowest zero-shot correlation.

\paragraph*{Improved ClinVar pathogenicity classification.} We analysed 114,221 variants from 11,298 proteins, calculating the area under the receiver operator characteristic curve (auROC) between predicted log-odds ratios and ClinVar labels. Fine-tuning increased auROC from 0.891 to 0.902 (+1.23\%, Figure \ref{fig:results}A). As seen in Figure \ref{fig:extendedbenchmark}B, after fine-tuning, log-odds ratio scores for pathogenic variants became more negative, and benign scores more positive, resulting in improved pathogenic-benign variant separation. We further used the balanced ClinVar subset with proteins having at least 10 benign and 10 pathogenic labels to examine the fine-tuning effect on individual proteins. As shown in Figure \ref{fig:results}D, fine-tuning consistently improved variant classification for most proteins. Notably, the improvements were larger for proteins with lower baseline auROCs (below 0.8), while proteins with strong zero-shot performance saw minimal gains. This analysis revealed a micro-averaged auROC increase from 0.860 to 0.877 (+1.98\%), and a macro-averaged auROC rise across proteins from 0.880 to 0.891 (+1.25\%).



\paragraph*{NLR fine-tuning improves other ESM models.} We further assessed the impact of NLR fine-tuning with ESM-1b and ESM-2. ESM-1b shares the same architecture as ESM-1v but was trained on different corpora and not directly optimised for variant effect predictions \citep{rives2021biological}. ESM-2 introduced improvements in architecture, number of training parameters and pre-training data, outperforming previous ESM models on protein structure prediction \citep{lin2023evolutionary}. Across all benchmarks, ESM-1b and ESM-2 exhibited performance improvements after NLR fine-tuning (Table \ref{tab:allperfs}). Notably, performance gains were higher for ESM-1b and ESM-2 compared to the ensemble of ESM-1v models on our DMS benchmarks, with up to 25.6\% increase in average Spearman correlation on ProteinGym for ESM-1b. This indicates that NLR's benefits extend beyond specific architectures and may be more significant in single-model settings. 

\begin{table}[h]
   \caption{Performance of ESM models in each evaluation benchmark before (Zero-Shot) and after NLR fine-tuning (+ NLR ft).}
   \label{tab:allperfs}
   \begin{center}
   \begin{tabular}{c|c|ccc}
     \toprule
      &  & MaveDB& ProteinGym & ClinVar \\
      &  & (avg Spearman)& (avg Spearman) & (auROC) \\
     \midrule
     ESM-1v      & Zero-Shot & 0.478	      & 0.331 	         & 0.891         \\
     ensemble    & + NLR ft  & \textbf{0.503}	& \textbf{0.396}	& \textbf{0.902} \\
     \midrule
     ESM-1b      & Zero-Shot & 0.451 	      & 0.309 	         & 0.913         \\
                 & + NLR ft  & \textbf{0.498}& \textbf{0.388}	& \textbf{0.919} \\
      \midrule
     ESM-2 (650M) & Zero-Shot & 0.482	      & 0.317	         & 0.884         \\
                 & + NLR ft  & \textbf{0.509}	& \textbf{0.393}	& \textbf{0.894} \\
     \bottomrule
   \end{tabular}
   \end{center}
\end{table}

\section{Discussion}
We have presented a novel approach to enhance variant effect predictions from PLMs by fine-tuning on DMS datasets. We applied score normalisation across DMS assays to tackle challenges in data integration and introduced a novel lightweight fine-tuning Normalised Log-odds Ratio head that allows PLM’s to efficiently learn from DMS data without adding task-specific parameters. After NLR fine-tuning, we observed moderate but consistent improvements across variant effect prediction benchmarks, proteins and ESM models, highlighting the robustness of the approach.
Despite fine-tuning with DMS data from a limited set of 25 proteins, we observed accuracy improvements in the classification of the clinical significance of variants from 11,298 proteins in ClinVar, and in the correlation with experimental measurements from independent DMS assays in ProteinGym and MaveDB. The improvements were more pronounced for proteins with a lower zero-shot performance and lesser representation in the pre-training datasets (for example, viral proteins). These findings demonstrate that our fine-tuning approach improves the performance of PLMs beyond self-supervision on naturally selected protein sequences.

This study focused on the ESM family of PLMs, which utilises single protein sequences as input. Other models like MSA Transformer, EVE, and AlphaMissense \citep{rao2021msa, frazer2021disease, cheng2023accurate} leverage Multiple Sequence Alignments (MSAs) in the input space. Alignment-based PLMs can perform better in proteins with a limited number of homologs in protein sequence databases, such as viral proteins, overcoming some of the performance gaps of ESM models. Notably, AlphaMissense has recently achieved SOTA performance on multiple missense variant effect prediction tasks, with reported ClinVar accuracy and DMS correlations superior to the NLR fine-tuned models in this study. Adapting NLR fine-tuning to MSA-based PLMs is a promising avenue for future work. Although we used DMS data from a limited set of 25 proteins, we observed that NLR fine-tuning can continue to scale model performance with more DMS data. We believe that NLR fine-tuning provides a lightweight and efficient approach to improve PLMs variant effect predictions as the volume, quality, and standards of DMS data continue to grow.

\end{spacing}



\clearpage

\bibliography{iclr2024_conference}

\begin{thebibliography}{32}
\providecommand{\natexlab}[1]{#1}
\providecommand{\url}[1]{\texttt{#1}}
\expandafter\ifx\csname urlstyle\endcsname\relax
  \providecommand{\doi}[1]{doi: #1}\else
  \providecommand{\doi}{doi: \begingroup \urlstyle{rm}\Url}\fi

\bibitem[Backman et~al.(2021)Backman, Li, Marcketta, Sun, Mbatchou, Kessler,
  Benner, Liu, Locke, Balasubramanian, et~al.]{backman2021exome}
Joshua~D Backman, Alexander~H Li, Anthony Marcketta, Dylan Sun, Joelle
  Mbatchou, Michael~D Kessler, Christian Benner, Daren Liu, Adam~E Locke,
  Suganthi Balasubramanian, et~al.
\newblock {Exome sequencing and analysis of 454,787 UK Biobank participants}.
\newblock \emph{Nature}, 599\penalty0 (7886):\penalty0 628--634, 2021.

\bibitem[Brandes et~al.(2023)Brandes, Goldman, Wang, Ye, and
  Ntranos]{brandes2023genome}
Nadav Brandes, Grant Goldman, Charlotte~H Wang, Chun~Jimmie Ye, and Vasilis
  Ntranos.
\newblock {Genome-wide prediction of disease variant effects with a deep
  protein language model}.
\newblock \emph{Nature Genetics}, 55\penalty0 (9):\penalty0 1512--1522, 2023.

\bibitem[Cheng et~al.(2023)Cheng, Novati, Pan, Bycroft, {\v{Z}}emgulyt{\.e},
  Applebaum, Pritzel, Wong, Zielinski, Sargeant, et~al.]{cheng2023accurate}
Jun Cheng, Guido Novati, Joshua Pan, Clare Bycroft, Akvil{\.e}
  {\v{Z}}emgulyt{\.e}, Taylor Applebaum, Alexander Pritzel, Lai~Hong Wong,
  Michal Zielinski, Tobias Sargeant, et~al.
\newblock {Accurate proteome-wide missense variant effect prediction with
  AlphaMissense}.
\newblock \emph{Science}, 381\penalty0 (6664):\penalty0 eadg7492, 2023.

\bibitem[Dunham \& Beltrao(2021)Dunham and Beltrao]{dunham2021exploring}
Alistair~S Dunham and Pedro Beltrao.
\newblock {Exploring amino acid functions in a deep mutational landscape}.
\newblock \emph{Molecular systems biology}, 17\penalty0 (7):\penalty0 e10305,
  2021.

\bibitem[Esposito et~al.(2019)Esposito, Weile, Shendure, Starita, Papenfuss,
  Roth, Fowler, and Rubin]{esposito2019mavedb}
Daniel Esposito, Jochen Weile, Jay Shendure, Lea~M Starita, Anthony~T
  Papenfuss, Frederick~P Roth, Douglas~M Fowler, and Alan~F Rubin.
\newblock {MaveDB: an open-source platform to distribute and interpret data
  from multiplexed assays of variant effect}.
\newblock \emph{Genome biology}, 20:\penalty0 1--11, 2019.

\bibitem[Findlay et~al.(2018)Findlay, Daza, Martin, Zhang, Leith, Gasperini,
  Janizek, Huang, Starita, and Shendure]{findlay2018accurate}
Gregory~M Findlay, Riza~M Daza, Beth Martin, Melissa~D Zhang, Anh~P Leith,
  Molly Gasperini, Joseph~D Janizek, Xingfan Huang, Lea~M Starita, and Jay
  Shendure.
\newblock {Accurate classification of BRCA1 variants with saturation genome
  editing}.
\newblock \emph{Nature}, 562\penalty0 (7726):\penalty0 217--222, 2018.

\bibitem[Fowler et~al.(2021)Fowler, Hurles, Adams, Gloyn, Hahn, Marks, Neal,
  Roth, Rubin, Starita, et~al.]{fowler2021atlas}
DM~Fowler, M~Hurles, DJ~Adams, AL~Gloyn, WC~Hahn, DS~Marks, JT~Neal, F~Roth,
  AF~Rubin, LM~Starita, et~al.
\newblock {The Atlas of Variant Effects (AVE) Alliance: understanding genetic
  variation at nucleotide resolution}.
\newblock \emph{Zenodo}, 2021.

\bibitem[Fowler \& Fields(2014)Fowler and Fields]{fowler2014deep}
Douglas~M Fowler and Stanley Fields.
\newblock {Deep mutational scanning: a new style of protein science}.
\newblock \emph{Nature methods}, 11\penalty0 (8):\penalty0 801--807, 2014.

\bibitem[Frazer et~al.(2021)Frazer, Notin, Dias, Gomez, Min, Brock, Gal, and
  Marks]{frazer2021disease}
Jonathan Frazer, Pascal Notin, Mafalda Dias, Aidan Gomez, Joseph~K Min, Kelly
  Brock, Yarin Gal, and Debora~S Marks.
\newblock {Disease variant prediction with deep generative models of
  evolutionary data}.
\newblock \emph{Nature}, 599\penalty0 (7883):\penalty0 91--95, 2021.

\bibitem[Gao et~al.(2023)Gao, Hamp, Ede, Schraiber, McRae, Singer-Berk, Yang,
  Dietrich, Fiziev, Kuderna, et~al.]{gao2023landscape}
Hong Gao, Tobias Hamp, Jeffrey Ede, Joshua~G Schraiber, Jeremy McRae, Moriel
  Singer-Berk, Yanshen Yang, Anastasia~SD Dietrich, Petko~P Fiziev, Lukas~FK
  Kuderna, et~al.
\newblock {The landscape of tolerated genetic variation in humans and
  primates}.
\newblock \emph{Science}, 380\penalty0 (6648):\penalty0 eabn8153, 2023.

\bibitem[Hsu et~al.(2022)Hsu, Nisonoff, Fannjiang, and
  Listgarten]{hsu2022learning}
Chloe Hsu, Hunter Nisonoff, Clara Fannjiang, and Jennifer Listgarten.
\newblock {Learning protein fitness models from evolutionary and assay-labeled
  data}.
\newblock \emph{Nature biotechnology}, 40\penalty0 (7):\penalty0 1114--1122,
  2022.

\bibitem[Hu et~al.(2021)Hu, Shen, Wallis, Allen-Zhu, Li, Wang, Wang, and
  Chen]{hu2021lora}
Edward~J Hu, Yelong Shen, Phillip Wallis, Zeyuan Allen-Zhu, Yuanzhi Li, Shean
  Wang, Lu~Wang, and Weizhu Chen.
\newblock {LoRA: Low-rank adaptation of large language models}.
\newblock \emph{arXiv preprint arXiv:2106.09685}, 2021.

\bibitem[Jagota et~al.(2023)Jagota, Ye, Albors, Rastogi, Koehl, Ioannidis, and
  Song]{jagota2023cross}
Milind Jagota, Chengzhong Ye, Carlos Albors, Ruchir Rastogi, Antoine Koehl,
  Nilah Ioannidis, and Yun~S Song.
\newblock Cross-protein transfer learning substantially improves disease
  variant prediction.
\newblock \emph{Genome Biology}, 24\penalty0 (1):\penalty0 182, 2023.

\bibitem[Karczewski et~al.(2020)Karczewski, Francioli, Tiao, Cummings,
  Alf{\"o}ldi, Wang, Collins, Laricchia, Ganna, Birnbaum,
  et~al.]{karczewski2020mutational}
Konrad~J Karczewski, Laurent~C Francioli, Grace Tiao, Beryl~B Cummings, Jessica
  Alf{\"o}ldi, Qingbo Wang, Ryan~L Collins, Kristen~M Laricchia, Andrea Ganna,
  Daniel~P Birnbaum, et~al.
\newblock {The mutational constraint spectrum quantified from variation in
  141,456 humans}.
\newblock \emph{Nature}, 581\penalty0 (7809):\penalty0 434--443, 2020.

\bibitem[Landrum et~al.(2018)Landrum, Lee, Benson, Brown, Chao, Chitipiralla,
  Gu, Hart, Hoffman, Jang, et~al.]{landrum2018clinvar}
Melissa~J Landrum, Jennifer~M Lee, Mark Benson, Garth~R Brown, Chen Chao,
  Shanmuga Chitipiralla, Baoshan Gu, Jennifer Hart, Douglas Hoffman, Wonhee
  Jang, et~al.
\newblock {ClinVar: improving access to variant interpretations and supporting
  evidence}.
\newblock \emph{Nucleic acids research}, 46\penalty0 (D1):\penalty0
  D1062--D1067, 2018.

\bibitem[Lek et~al.(2016)Lek, Karczewski, Minikel, Samocha, Banks, Fennell,
  O'Donnell-Luria, Ware, Hill, Cummings, et~al.]{lek2016analysis}
Monkol Lek, Konrad~J Karczewski, Eric~V Minikel, Kaitlin~E Samocha, Eric Banks,
  Timothy Fennell, Anne~H O'Donnell-Luria, James~S Ware, Andrew~J Hill, Beryl~B
  Cummings, et~al.
\newblock {Analysis of protein-coding genetic variation in 60,706 humans}.
\newblock \emph{Nature}, 536\penalty0 (7616):\penalty0 285--291, 2016.

\bibitem[Lin et~al.(2023{\natexlab{a}})Lin, Wells, Wang, Orengo, and
  Martin]{lin2023varipred}
Weining Lin, Jude Wells, Zeyuan Wang, Christine Orengo, and Andrew~CR Martin.
\newblock {VariPred: Enhancing Pathogenicity Prediction of Missense Variants
  Using Protein Language Models}.
\newblock \emph{bioRxiv}, pp.\  2023--03, 2023{\natexlab{a}}.

\bibitem[Lin et~al.(2023{\natexlab{b}})Lin, Akin, Rao, Hie, Zhu, Lu, Smetanin,
  Verkuil, Kabeli, Shmueli, et~al.]{lin2023evolutionary}
Zeming Lin, Halil Akin, Roshan Rao, Brian Hie, Zhongkai Zhu, Wenting Lu, Nikita
  Smetanin, Robert Verkuil, Ori Kabeli, Yaniv Shmueli, et~al.
\newblock {Evolutionary-scale prediction of atomic-level protein structure with
  a language model}.
\newblock \emph{Science}, 379\penalty0 (6637):\penalty0 1123--1130,
  2023{\natexlab{b}}.

\bibitem[Livesey \& Marsh(2023)Livesey and Marsh]{livesey2023updated}
Benjamin~J Livesey and Joseph~A Marsh.
\newblock {Updated benchmarking of variant effect predictors using deep
  mutational scanning}.
\newblock \emph{Molecular Systems Biology}, pp.\  e11474, 2023.

\bibitem[Meier et~al.(2021)Meier, Rao, Verkuil, Liu, Sercu, and
  Rives]{meier2021language}
Joshua Meier, Roshan Rao, Robert Verkuil, Jason Liu, Tom Sercu, and Alex Rives.
\newblock {Language models enable zero-shot prediction of the effects of
  mutations on protein function}.
\newblock \emph{Advances in Neural Information Processing Systems},
  34:\penalty0 29287--29303, 2021.

\bibitem[Notin et~al.(2023)Notin, Kollasch, Ritter, van Niekerk, Paul, Spinner,
  Rollins, Shaw, Weitzman, Frazer, et~al.]{notin2023proteingym}
Pascal Notin, Aaron~W Kollasch, Daniel Ritter, Lood van Niekerk, Steffanie
  Paul, Hansen Spinner, Nathan Rollins, Ada Shaw, Ruben Weitzman, Jonathan
  Frazer, et~al.
\newblock {ProteinGym: Large-Scale Benchmarks for Protein Design and Fitness
  Prediction}.
\newblock \emph{bioRxiv}, pp.\  2023--12, 2023.

\bibitem[Radford et~al.(2023)Radford, Tan, Andersson, Stephenson, Gardner,
  Ironfield, Waters, Gitterman, Lindsay, Abascal,
  et~al.]{radford2023saturation}
Elizabeth~J Radford, Hong-Kee Tan, Malin~HL Andersson, James~D Stephenson,
  Eugene~J Gardner, Holly Ironfield, Andrew~J Waters, Daniel Gitterman, Sarah
  Lindsay, Federico Abascal, et~al.
\newblock {Saturation genome editing of DDX3X clarifies pathogenicity of
  germline and somatic variation}.
\newblock \emph{Nature Communications}, 14\penalty0 (1):\penalty0 7702, 2023.

\bibitem[Rao et~al.(2021)Rao, Liu, Verkuil, Meier, Canny, Abbeel, Sercu, and
  Rives]{rao2021msa}
Roshan~M Rao, Jason Liu, Robert Verkuil, Joshua Meier, John Canny, Pieter
  Abbeel, Tom Sercu, and Alexander Rives.
\newblock {MSA transformer}.
\newblock In \emph{{International Conference on Machine Learning}}, pp.\
  8844--8856. PMLR, 2021.

\bibitem[Rives et~al.(2021)Rives, Meier, Sercu, Goyal, Lin, Liu, Guo, Ott,
  Zitnick, Ma, et~al.]{rives2021biological}
Alexander Rives, Joshua Meier, Tom Sercu, Siddharth Goyal, Zeming Lin, Jason
  Liu, Demi Guo, Myle Ott, C~Lawrence Zitnick, Jerry Ma, et~al.
\newblock {Biological structure and function emerge from scaling unsupervised
  learning to 250 million protein sequences}.
\newblock \emph{Proceedings of the National Academy of Sciences}, 118\penalty0
  (15):\penalty0 e2016239118, 2021.

\bibitem[Rubin et~al.(2021)Rubin, Min, Rollins, Da, Esposito, Harrington,
  Stone, Bianchi, Dias, Frazer, et~al.]{rubin2021mavedb}
Alan~F Rubin, Joseph~K Min, Nathan~J Rollins, Estelle~Y Da, Daniel Esposito,
  Matthew Harrington, Jeremy Stone, Aisha~Haley Bianchi, Mafalda Dias, Jonathan
  Frazer, et~al.
\newblock {MaveDB v2: a curated community database with over three million
  variant effects from multiplexed functional assays}.
\newblock \emph{bioRxiv}, pp.\  2021--11, 2021.

\bibitem[Schmirler et~al.(2023)Schmirler, Heinzinger, and
  Rost]{schmirler2023fine}
Robert Schmirler, Michael Heinzinger, and Burkhard Rost.
\newblock {Fine-tuning protein language models boosts predictions across
  diverse tasks}.
\newblock \emph{bioRxiv}, pp.\  2023--12, 2023.

\bibitem[Sledzieski et~al.(2023)Sledzieski, Kshirsagar, Baek, Berger, Dodhia,
  and Ferres]{sledzieski2023democratizing}
Samuel Sledzieski, Meghana Kshirsagar, Minkyung Baek, Bonnie Berger, Rahul
  Dodhia, and Juan~Lavista Ferres.
\newblock {Democratizing Protein Language Models with Parameter-Efficient
  Fine-Tuning}.
\newblock \emph{bioRxiv}, pp.\  2023--11, 2023.

\bibitem[Steinegger \& S{\"o}ding(2017)Steinegger and
  S{\"o}ding]{steinegger2017mmseqs2}
Martin Steinegger and Johannes S{\"o}ding.
\newblock {MMseqs2 enables sensitive protein sequence searching for the
  analysis of massive data sets}.
\newblock \emph{Nature biotechnology}, 35\penalty0 (11):\penalty0 1026--1028,
  2017.

\bibitem[Ulyanov et~al.(2016)Ulyanov, Vedaldi, and
  Lempitsky]{ulyanov2016instance}
Dmitry Ulyanov, Andrea Vedaldi, and Victor Lempitsky.
\newblock {Instance normalization: The missing ingredient for fast
  stylization}.
\newblock \emph{arXiv preprint arXiv:1607.08022}, 2016.

\bibitem[Umerenkov et~al.(2023)Umerenkov, Nikolaev, Shashkova, Strashnov,
  Sindeeva, Shevtsov, Ivanisenko, and Kardymon]{umerenkov2023prostata}
Dmitriy Umerenkov, Fedor Nikolaev, Tatiana~I Shashkova, Pavel~V Strashnov,
  Maria Sindeeva, Andrey Shevtsov, Nikita~V Ivanisenko, and Olga~L Kardymon.
\newblock {PROSTATA: a framework for protein stability assessment using
  transformers}.
\newblock \emph{Bioinformatics}, 39\penalty0 (11):\penalty0 btad671, 2023.

\bibitem[{UniProt Consortium}(2023)]{uniprot2023uniprot}
{UniProt Consortium}.
\newblock {UniProt: the universal protein knowledgebase in 2023}.
\newblock \emph{Nucleic Acids Research}, 51\penalty0 (D1):\penalty0 D523--D531,
  2023.

\bibitem[{wwPDB Consortium}(2019)]{wwpdb2019protein}
{wwPDB Consortium}.
\newblock {Protein Data Bank: the single global archive for 3D macromolecular
  structure data}.
\newblock \emph{Nucleic acids research}, 47\penalty0 (D1):\penalty0 D520--D528,
  2019.

\end{thebibliography}
\bibliographystyle{iclr2024_conference}

\clearpage

\appendix
\section{Appendix}

\setcounter{equation}{0}
\setcounter{figure}{0}
\setcounter{table}{0}
\renewcommand{\theequation}{S\arabic{equation}}
\renewcommand{\thefigure}{S\arabic{figure}}
\renewcommand{\thetable}{S\arabic{table}}
\renewcommand{\bibnumfmt}[1]{[S#1]}
\renewcommand{\citenumfont}[1]{S#1}



\subsection{Fine-tuning details}
\label{subsec:finetuning}

\paragraph*{Experimental setup.} The base model for hyperparameter search and ablations was the first ESM-1v model checkpoint (esm1v\_t33\_650M\_UR90S\_1). Experiments ran for a maximum of 10 epochs, saving checkpoints based on peak micro-averaged Spearman correlation on the validation fold. We used an effective batch size of 128 samples across 8 V100s (32GB), training in Distributed Data Parallel mode, with two samples per device and accumulating gradients for 8 steps. All experiments performed learning rate warmup from 0 to the peak learning rate over 200 steps, and the peak learning rate was chosen via grid search (1e-5, 2e-5, 5e-5, and 1e-4). Cross-validation experiments revealed an optimal learning rate of 2e-5, exhibiting overfitting after 2,000 optimization steps. 

\paragraph*{Cross-validation.} Five-fold cross-validation was performed on the training set (25 proteins from MaveDB) to compare model architectures and hyperparameter settings. DMS data for 5 proteins was kept in each validation fold while 20 proteins were used for model training. The cross-validation scheme grouped variant instances by protein cluster to assess generalisation to unseen and dissimilar proteins. In each run, the absolute improvement in Spearman correlation over the zero-shot model was computed at each step ($t$): 

\begin{equation}
   improvement(t) = Spearman(t) - Spearman(t=0)
\end{equation}

where $Spearman(t$=$0)$ represents the correlation before starting model fine-tuning and $Spearman(t)$ represents the correlation after $t$ optimization steps. We save the model checkpoint at the step ($t_{max}$) with peak Spearman improvement.

\paragraph*{Log-odds ratio matrix normalisation.} During NLR fine-tuning, the log-odds ratio matrix is computed for each training instance, and min-max normalisation is then independently applied to each matrix:
\begin{equation}
   X_{scaled} = \frac{X-X_{min}}{X_{max}-X_{min}} (range_{max}-range_{min}) + range_{min}
 \end{equation}
 where $X$ is the log-odds ratio matrix, $X_{min}$ and $X_{max}$ are its minimum and maximum values (masking out padded positions), and $range_{min}$ and $range_{max}$ define the desired normalised range, -2 and 2, respectively, aligning with DMS scores. Normalisation was only performed during fine-tuning but not at inference time. 

\subsection{Regularisation Experiments} 

Training instances are defined by the wildtype protein sequence and a single DMS variant score. Each DMS variant score corresponds to a specific amino acid substitution at a certain position, and its corresponding prediction is sampled from the log-odds ratio matrix. Given the large number of DMS scores per protein sequence, the model is repeatedly exposed to the same input sequence during training which could make it prone to overfitting. To test for this, we explored the effect of two regularisation strategies: input masking and layer freezing. 

\subsubsection{Input masking}
\label{subsec:masking}
Randomly masking a percentage of input tokens for each variant instance has the potential to prevent the model relying on local, noise-based correlations between the amino acid sequence and the DMS output distribution. To test this we experimented with masking 5, 15 and 30\% of input tokens for each instance during training, while at inference time, only the variant position was masked and the remaining context was shown to the model. Despite the potential for regularisation, no positive effects were observed when masking the input sequences (see Table \ref{tab:maskedtokens}). This result may be attributed to (1) a potential mismatch between the training and validation input distribution, since masked tokens are seen in a higher proportion during training than inference, when only the variant position was masked, and (2) strong dependencies between the effects of single nucleotide variants and specific motifs within the context of those variants. The latter scenario could result in masking of crucial information during training that prevents effective learning. 

\begin{table}[h]
   \caption{Peak Spearman correlation improvement across folds in MaveDB (mean $\pm$ std) when randomly masking input tokens.}
   \label{tab:maskedtokens}
   \begin{center}
   \begin{tabular}{cc}
     \toprule
      Inputed tokens masked (\%) & $improvement(t_{max})$ \\
     \midrule
     0      & 0.033 $\pm$ 0.008       \\
     5      & 0.022 $\pm$ 0.009       \\
     15     & 0.019 $\pm$ 0.004       \\
     30     & 0.017 $\pm$ 0.007       \\

     \bottomrule
   \end{tabular}
   \end{center}
\end{table}

\begin{spacing}{0.95}
\subsubsection{Layer freezing}
\label{subsec:freezing}
Layer freezing has the potential to regularise the model by preserving the information learned during pre-training in the frozen layers while fine-tuning the remaining encoder blocks. However, while progressively freezing ESM-1v we did not observe any clear improvements (see Table \ref{tab:freezing}). It is worth noting that the high variance across folds might hinder more general conclusions about the potential optimal number of layers frozen. However, three important conclusions were taken from this analysis:
\begin{enumerate}
  \item As a result of the experiment freezing 33 layers, it was clear that tuning transformer encoder blocks is essential to obtain higher improvements on Spearman correlation, and exclusively tuning the language modelling head only provides a minor improvement.
  \item Only unfreezing one encoder block performed worse than the rest of the experiments, suggesting that modelling this task requires a higher number of trainable parameters and learning higher-order token interactions. 
  \item From the results above, it was not clear that any number of frozen transformer encoder block experiments outperformed the fully unfrozen model. However, future experiments with parameter-efficient fine-tuning strategies \citep{hu2021lora} could reveal an optimal number of parameters to tune for this task.
\end{enumerate}

\begin{table}[h]
   \caption{Peak Spearman correlation improvement across folds in MaveDB (mean $\pm$std) when freezing layers.}
   \label{tab:freezing}
   \begin{center}
   \begin{tabular}{cc}
     \toprule
      Layers frozen & $improvement(t_{max})$ \\
     \midrule
     0       & 0.033 $\pm$ 0.008       \\
     12      & 0.027 $\pm$ 0.015       \\
     16      & 0.027 $\pm$ 0.014       \\
     20      & 0.027 $\pm$ 0.014       \\
     24      & 0.028 $\pm$ 0.012       \\
     28      & 0.031 $\pm$ 0.015       \\
     32      & 0.022 $\pm$ 0.019       \\
     33$^\dagger$     & 0.001 $\pm$ 0.007       \\
     \bottomrule
   \end{tabular} \\
   $^\dagger$ Only the Language Modelling head was fine-tuned.
   \end{center}

\end{table}

\subsection{Data Scaling}
\label{subsec:datascale}

We conducted experiments to evaluate how performance improvement scales with the number of training proteins that have DMS data. Analogous to regularisation experiments, we performed five-fold cross-validation on the training set. However, in each run, we randomly selected a subset of $n$ proteins from the training folds, and conducted experiments with $n$ being 2, 5, 10, 15 or 20 and ran all the experiments for five epochs. 
The results, which are displayed in Figure \ref{fig:datascaling}, demonstrate that the improvements of NLR fine-tuning scale with the number of proteins with DMS data. Furthemore, when increasing the amount of DMS data, we observed a consistent upward trend in improvement across all folds. 
Overall, these results indicate that NLR fine-tuning can effectively scale the performance of Protein Language Models (PLMs) in variant effect prediction as the number of DMS assays openly available contines to grow. 
\end{spacing}

\clearpage

\subsection{Tables and Figures}

\begin{table}[h]
   \caption{List of proteins and DMS assays from MaveDB used in this study.}
   \begin{center}
   \includegraphics[width=1.\textwidth]{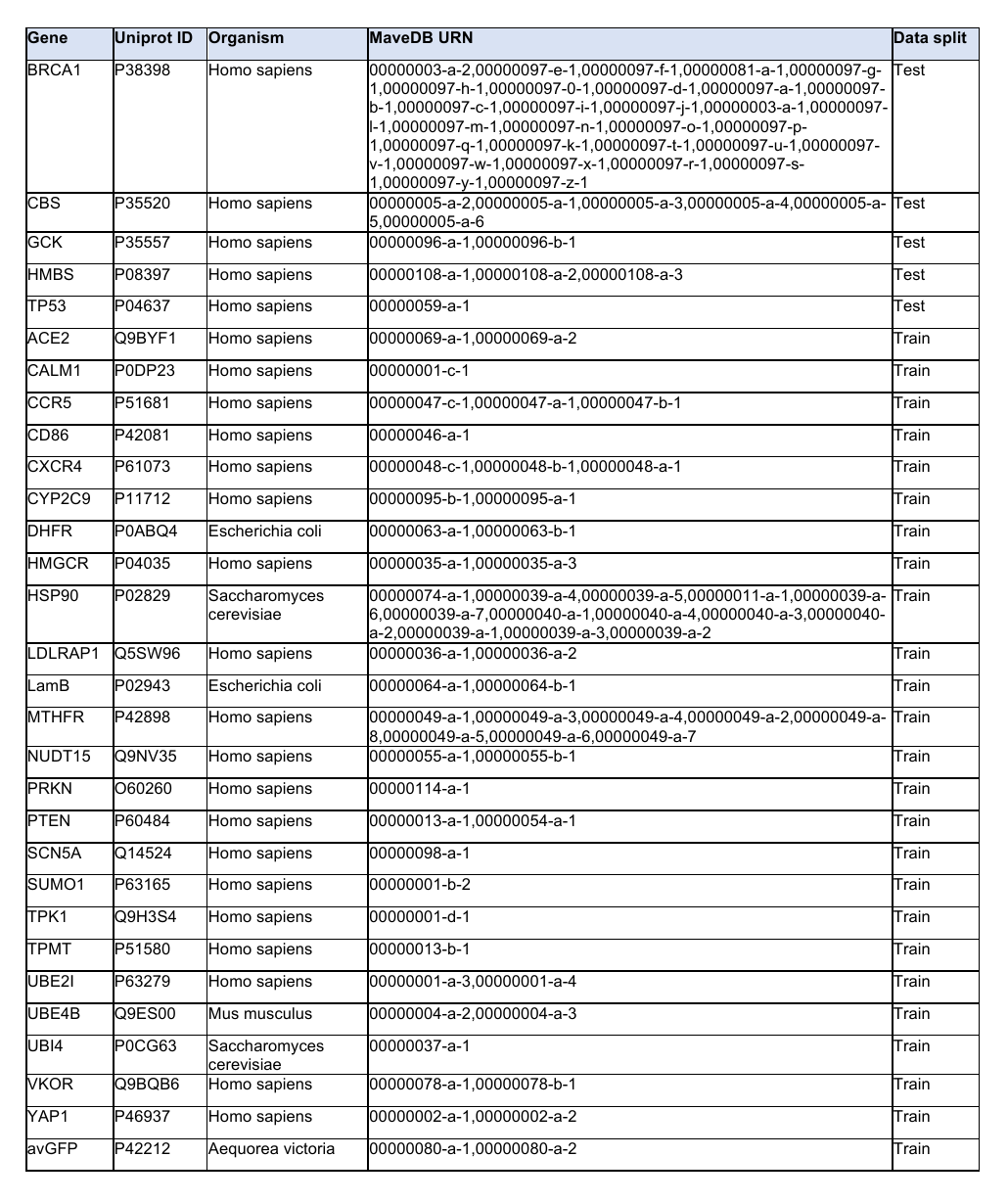}
\end{center}
   
     \label{tab:protlist}
\end{table}

\begin{table}[h]
   \caption{Micro-averaged Spearman correlation across MaveDB test proteins stratified by 
   ESM-1v model versions. Improvement is reported as \% of increase over Zero-Shot baseline.}
   \label{tab:esm1vs}
   \begin{center}
   \begin{tabular}{c|ccc}
     \toprule
     \textbf{ESM-1v checkpoint } & \textbf{Zero-Shot} & \textbf{NLR fine-tuned}& \textbf{\% improvement} \\
     \midrule
     1 & 0.444 & \textbf{0.476}  & 7.2 \\
     2 & 0.438 & \textbf{0.479}  & 9.4 \\
     3 & 0.456 & \textbf{0.481}  & 5.5 \\
     4 & 0.456 & \textbf{0.487}  & 6.8 \\
     5 & 0.452 & \textbf{0.490}  & 8.4 \\
     Ensemble 1-5   & 0.478 & \textbf{0.503}  & 5.2 \\
     \bottomrule
   \end{tabular}
   \end{center}
   \end{table}

\begin{figure}[h]
   \begin{center}
   \includegraphics[width=0.9\textwidth]{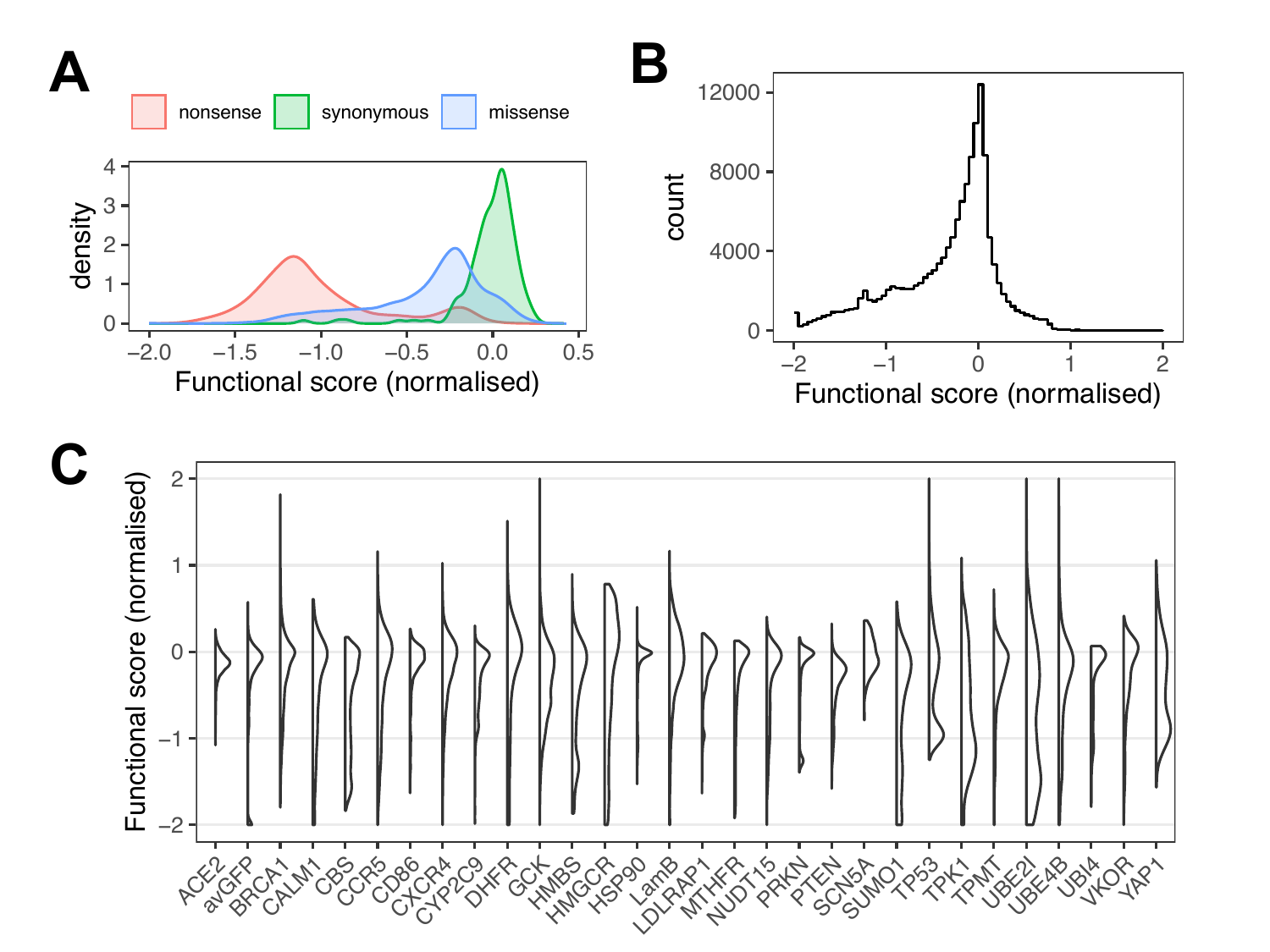}
\end{center}
   \caption{Normalisation of DMS functional scores. 
   A) Distribution of DMS functional scores for missense, nonsense, 
   and synonymous variants in MaveDB for the PTEN protein after score 
   rescaling and normalisation, shown as density functions.
   B) Stacked histogram of DMS functional scores for 
   all missense variants in the 30 proteins of the final normalised MaveDB dataset.
   C) Distributions of normalised DMS functional scores for each of the 30 proteins 
   in the MaveDB dataset, shown as half-violin plots.
    }
     \label{fig:mavedb}
\end{figure}

\begin{figure}[h]
   \begin{center}
   \includegraphics[width=1.\textwidth]{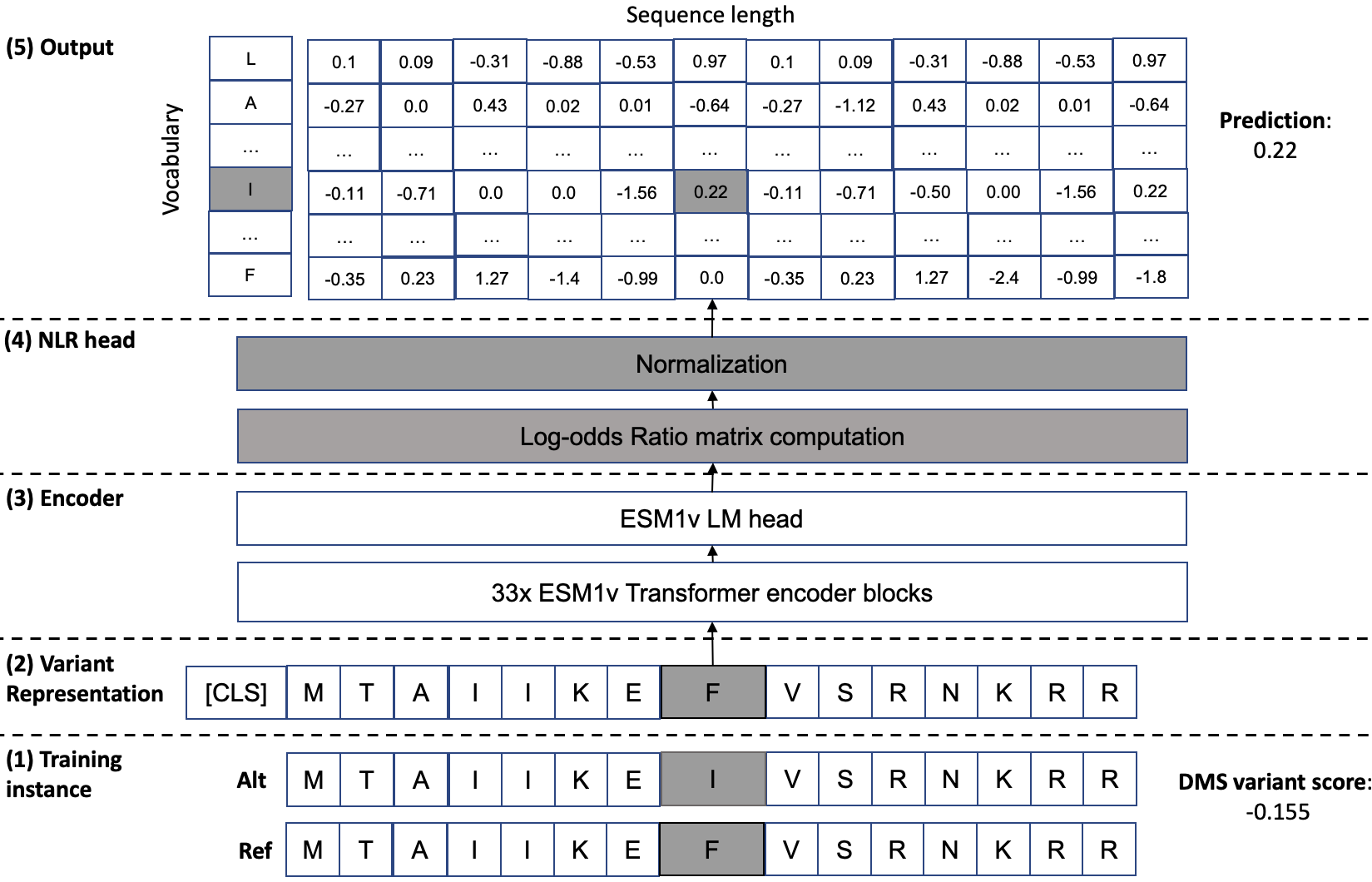}
\end{center}
   \caption{Diagram of the fine-tuning ESM-1v architecture with Normalised Log-odds Ratio (NLR) head. 
   (1) Training instance with a single amino acid swap and its DMS label. 
   (2) Input token representation for the wild-type sequence. 
   (3) ESM-1v pre-trained encoder blocks + Language Modelling head. 
   (4) Fine-tuning blocks, including log-odds ratio matrix calculation and normalisation layers. 
   (5) Output matrix with the chosen cell reflecting the predicted score for the input variant.}
     \label{fig:finetuning}
\end{figure}

\begin{figure}[h]
   \begin{center}
   \includegraphics[width=1.\textwidth]{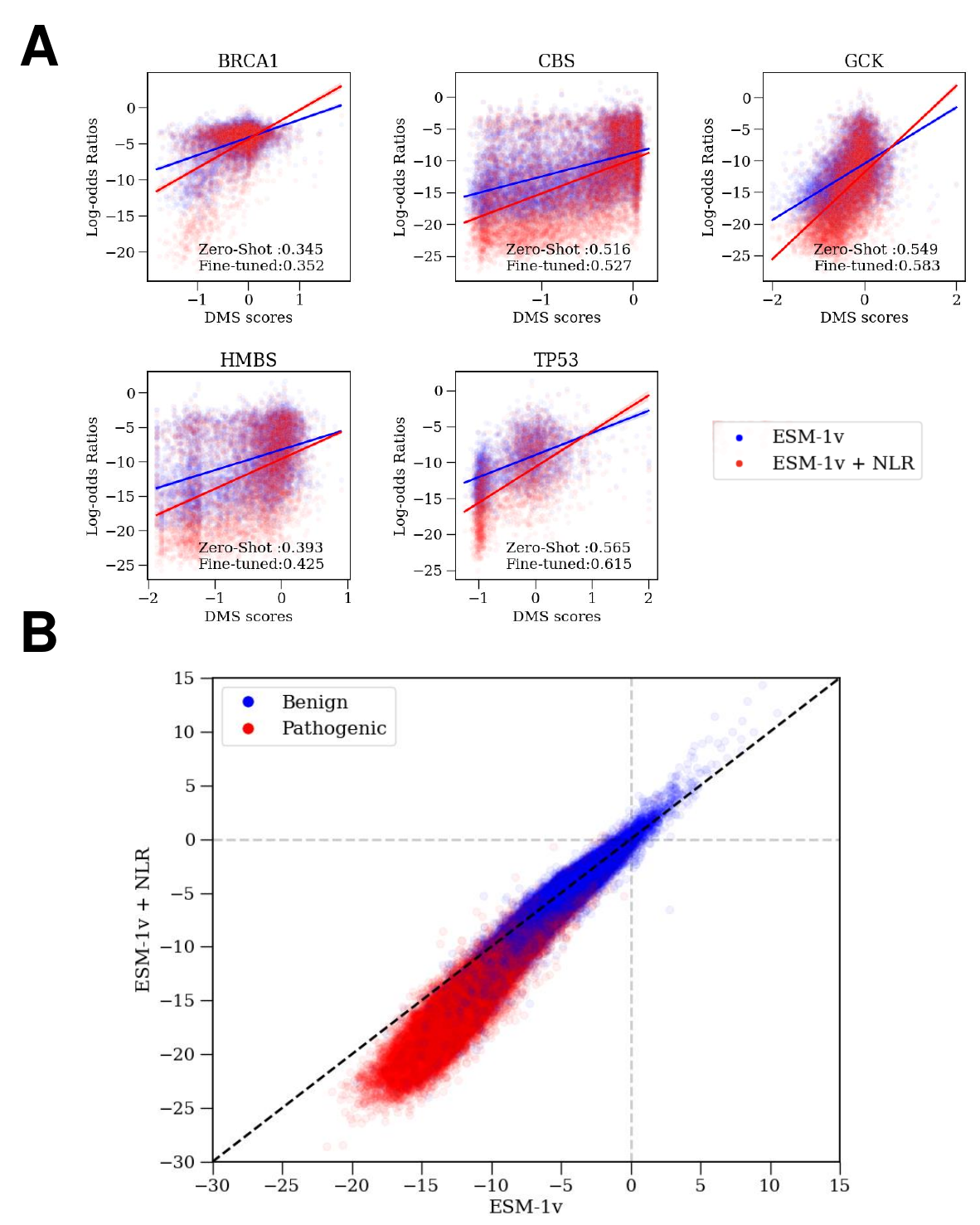}
\end{center}
   \caption{A) DMS vs. predicted log-odds ratios for each variant in the MaveDB test set, stratified by gene. In blue are the scores from the zero-shot ESM-1v, while in red are the NLR fine-tuned scores. A linear fit is displayed together with the score distribution and the Spearman correlation is shown for each distribution. B) Predicted log-odds ratios from ESM-1v vs. NLR-finetuned ESM-1v, for each variant in the ClinVar benchmark. In blue and red are the benign and pathogenic variants, respectively.}
     \label{fig:extendedbenchmark}
\end{figure}

\begin{figure}[h]
   \begin{center}
   \includegraphics[width=1.\textwidth]{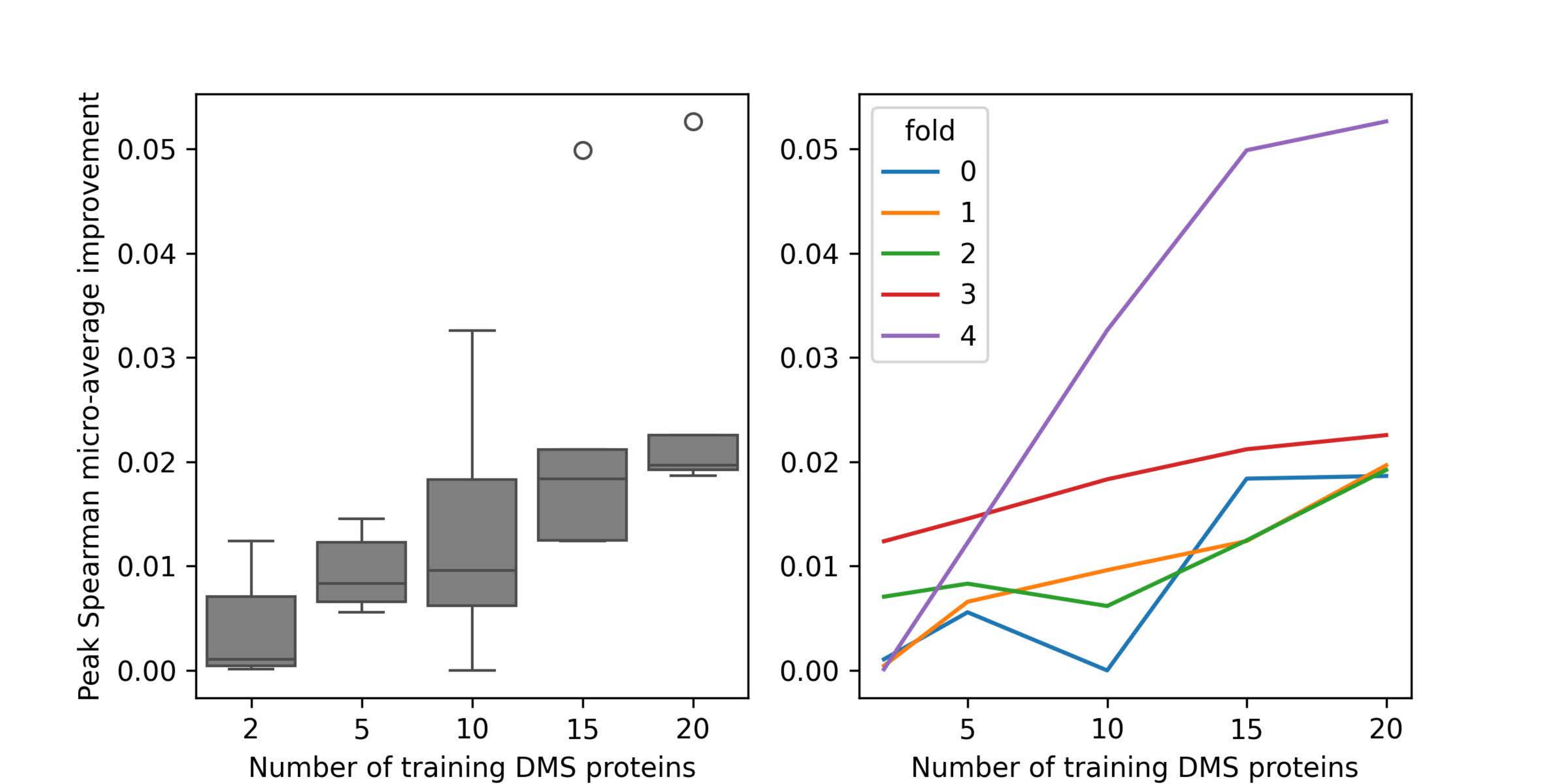}
\end{center}
   \caption{Peak Spearman correlation improvement across validation folds in MaveDB as a function of the number of training proteins. Left-side panel exhibits the improvement as a box-plot distribution across validation folds. The right-side panel shows the improvement stratified by validation fold.}
     \label{fig:datascaling}
\end{figure}

\end{document}